\documentstyle[epsfig]{aprim10}
\input{epsf}

\newif\ifAMStwofonts

\ifoldfss
  \ifCUPmtlplainloaded \else
    \NewTextAlphabet{textbfit} {cmbxti10} {}
    \NewTextAlphabet{textbfss} {cmssbx10} {}
    \NewMathAlphabet{mathbfit} {cmbxti10} {} 
    \NewMathAlphabet{mathbfss} {cmssbx10} {} 
  \fi
  \ifAMStwofonts
    \ifCUPmtlplainloaded \else
      \NewSymbolFont{upmath} {eurm10}
      \NewSymbolFont{AMSa} {msam10}
      \NewMathSymbol{\upi}     {0}{upmath}{19}
      \NewMathSymbol{\umu}     {0}{upmath}{16}
      \NewMathSymbol{\upartial}{0}{upmath}{40}
      \NewMathSymbol{\leqslant}{3}{AMSa}{36}
      \NewMathSymbol{\geqslant}{3}{AMSa}{3E}

       \let\le=\leqslant
       
    \fi
  \fi
\fi 

\ifnfssone
  \newmathalphabet{\mathit}
  \addtoversion{normal}{\mathit}{cmr}{m}{it}
  \addtoversion{bold}{\mathit}{cmr}{bx}{it}

\newcommand{\la}{\mbox{\raisebox{-0.6ex}{$\stackrel{\textstyle<}{\sim}$}}}
\newcommand{\ga}{\mbox{\raisebox{-0.6ex}{$\stackrel{\textstyle>}{\sim}$}}}
  \newmathalphabet{\mathbfit} 
  \addtoversion{normal}{\mathbfit}{cmr}{bx}{it}
  \addtoversion{bold}{\mathbfit}{cmr}{bx}{it}
  \newmathalphabet{\mathbfss} 
  \addtoversion{normal}{\mathbfss}{cmss}{bx}{n}
  \addtoversion{bold}{\mathbfss}{cmss}{bx}{n}
  \ifAMStwofonts
    \ifCUPmtlplainloaded \else
      \UseAMStwoboldmath
      \makeatletter
      \new@mathgroup\upmath@group
      \define@mathgroup\mv@normal\upmath@group{eur}{m}{n}
      \define@mathgroup\mv@bold\upmath@group{eur}{b}{n}
      \edef\UPM{\hexnumber\upmath@group}
      \new@mathgroup\amsa@group
      \define@mathgroup\mv@normal\amsa@group{msa}{m}{n}
      \define@mathgroup\mv@bold\amsa@group{msa}{m}{n}
      \edef\AMSa{\hexnumber\amsa@group}
      \makeatother
      \mathchardef\upi="0\UPM19
      \mathchardef\umu="0\UPM16
      \mathchardef\upartial="0\UPM40
      \mathchardef\leqslant="3\AMSa36
      \mathchardef\geqslant="3\AMSa3E

       \let\le=\leqslant

    \fi
  \fi
\fi 

\ifnfsstwo
  \DeclareMathAlphabet{\mathbfit}{OT1}{cmr}{bx}{it}
  \SetMathAlphabet\mathbfit{bold}{OT1}{cmr}{bx}{it}
  \DeclareMathAlphabet{\mathbfss}{OT1}{cmss}{bx}{n}
  \SetMathAlphabet\mathbfss{bold}{OT1}{cmss}{bx}{n}
  \ifAMStwofonts
    \ifCUPmtlplainloaded \else
      \DeclareSymbolFont{UPM}{U}{eur}{m}{n}
      \SetSymbolFont{UPM}{bold}{U}{eur}{b}{n}
      \DeclareSymbolFont{AMSa}{U}{msa}{m}{n}
      \DeclareMathSymbol{\upi}{0}{UPM}{"19}
      \DeclareMathSymbol{\umu}{0}{UPM}{"16}
      \DeclareMathSymbol{\upartial}{0}{UPM}{"40}
      \DeclareMathSymbol{\leqslant}{3}{AMSa}{"36}
      \DeclareMathSymbol{\geqslant}{3}{AMSa}{"3E}

       \let\le=\leqslant

    \fi
  \fi
\fi 

\ifCUPmtlplainloaded \else
  \ifAMStwofonts \else 
    \def\upi{\pi}
    \def\umu{\mu}
    \def\upartial{\partial}
  \fi
\fi

\title[How rapidly do neutron stars spin at birth?]{How rapidly do neutron stars spin at birth?}

\author[Soria et al.]
       {R. Soria$^1$, R. Perna$^2$, D. Pooley$^3$, and L. Stella$^4$\\
        $^1$Mullard Space Science Laboratory, University College London, 
Holmbury St Mary, Dorking RH5 6NT, UK\\
        $^2$JILA and Department of Astrophysical and Planetary Sciences, 
University of Colorado, Boulder, CO 80309, USA\\
        $^3$Astronomy Department, University of Wisconsin, Madison, WI 53706, USA\\ 
        $^4$INAF - Osservatorio Astronomico di Roma, Via Frascati 33, I-00040 Rome, Italy}
\date{}

\pagerange{\pageref{firstpage}--\pageref{lastpage}}
\pubyear{2008}

\begin{document}

\maketitle

\label{firstpage}

\begin{abstract}
We have studied the X-ray properties of ageing historical 
core-collapse supernovae in nearby galaxies, 
using archival data from {\it Chandra}, 
{\it XMM-Newton} and {\it Swift}.
We found possible evidence of a young X-ray pulsar in SN\,1968D 
and in few other sources, but none more luminous than $\sim$ a few 
$10^{37}$ erg s$^{-1}$. We compared the observational limits 
to the X-ray pulsar luminosity 
distribution with the results of Monte Carlo simulations 
for a range of birth parameters. We conclude that a pulsar population 
dominated by periods $\la 40$ ms at birth is ruled out by the data.
\end{abstract}


\section{X-ray supernovae and pulsars}

X-ray imaging missions  
have detected $\approx 40$ young core-collapse supernovae (SNe); 
this has led to a deeper understanding of the initial 
phases of shock expansion. However, only $\approx 10$ SNe have been detected  
in the X-ray band at ages $> 20$ yr, even though there is 
a reliable optical record of historical SNe in nearby galaxies 
over $\approx 100$ yrs. On the other hand, 
core-collapse SN remnants are found with much older ages 
($\ga 350$ yrs). But we still know little 
about the shock propagation into the ambient medium 
during the transition between the SN and SN remnant 
phases (ages of $\sim 30$--$300$ yrs).  
Thus, it is important to search for and study historical 
SNe as far as possible into this age range.

From the X-ray point of view, the main phases of 
SN evolution are \cite{fra96}:\\
\noindent
a) {\it SN phase = circumstellar medium interaction}: the shock 
propagates into the stellar wind of the progenitor star.
Emission from the forward shock (gas temperature $\ga$ a few keV) 
dominates only in the very early epoch ($t \la 100$ d); then, after 
the shell of ejecta becomes optically thin, the reverse 
shock provides the X-ray photons (thermal spectrum at 
$kT \approx 0.5$--$1$ keV). The older the SN, the further 
back in time we probe the mass-loss rate of the progenitor 
star, and hence its evolution in the last few thousand years.\\ 
b) {\it SN Remnant phase = interstellar medium interaction}: 
the expanding shock reaches the boundary 
between the stellar wind bubble and the surrounding 
interstellar medium. When that happens, the 
luminosity decay flattens out, and the remnant 
may keep a constant or very slowly declining luminosity 
and temperature for $\sim$ a few $10^3$ yrs. 

This scenario can have an additional component.
A large fraction of core-collapse SNe 
will leave behind a spinning, isolated neutron star (NS) 
powered by electromagnetic losses. 
For a dipole magnetic field $B$, the energy loss is given by (assuming
an orthogonal rotator) 
$\dot{E}_{\rm rot}= {B^2\,\Omega^4\,R^6}/{6c^3}$,
where $R$ is the NS radius and $\Omega$ its angular velocity. 
There is an empirical relation \cite{pos02} between the
X-ray luminosity of the pulsar and its rotational energy loss:  
$\log L_{\rm X,[2-10]}= 1.34\,\log \dot{E}_{\rm rot} -15.34$.
Analogous relations were inferred by Li et al. \shortcite{li08} 
and Kargaltsev \& Pavlov \shortcite{kar08}.
The X-ray luminosity includes both the magnetospheric emission from the
pulsar itself, and the synchrotron emission from the pulsar wind nebula. 
It has a power-law spectrum with photon index $\Gamma \approx 2$.
It declines as $L_{\rm X}=L_{\rm{X},0}(1+t/t_0)^{-2}$, where
$t_0\sim$ a few $10^3$ yrs.
For $t\la t_0$ the pulsar flux does not vary significantly. 
Therefore, for some SNe with very energetic pulsars (e.g., the Crab), 
the hard X-ray emission from the pulsar 
itself becomes progressively more important than 
the thermal emission from shocked gas, as the SN ages. 
In external galaxies, X-ray imaging cannot resolve the shocked gas from the point-like 
pulsar, but X-ray spectra and colors can tell which component dominates.

The pulsar luminosity distribution is determined 
by the magnetic field and initial spin period distributions. 
The magnetic field distribution is relatively well constrained 
($\sim 10^{12}$--$10^{13}$ Gauss). Instead, the birth spin 
is very uncertain. Some radio studies have suggested a large
population of millisecond pulsars \cite{arz02}, while
others have claimed millisecond pulsars are very rare \cite{fau06}.
We have recently proposed a new, independent method to constrain the
initial period distribution of pulsars and solve this controversy \cite{per08}.
Using {\it Chandra}, {\it Swift} and {\it XMM-Newton} data,
we have compared the observed X-ray luminosities (or
upper limits) of young SNe with the predicted distribution of pulsar
X-ray luminosities. 

\begin{figure}
\hspace{0.3cm}
\includegraphics[width=0.47\textwidth]{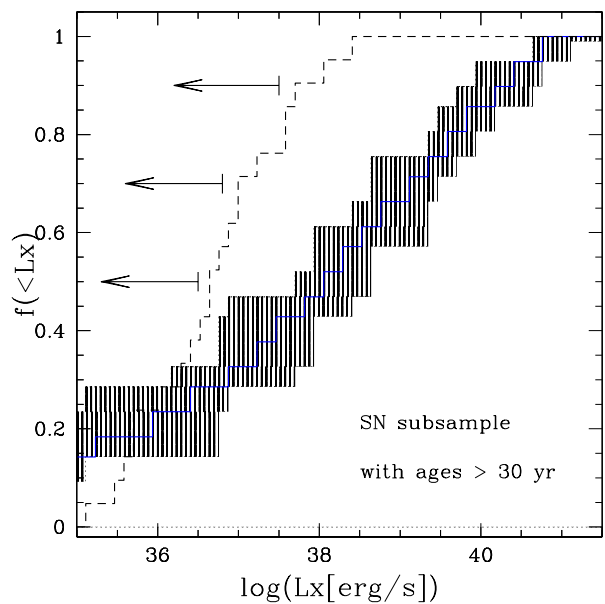}
\caption{\footnotesize  
The dashed line shows the distribution of $2$--$10$ keV 
luminosities (measurements or upper limits) 
for historical SNe with ages $> 30$ yrs; they are 
treated as upper limits for the underlying pulsar 
luminosities. The solid line (with shaded 1$\sigma$ error) 
shows the predicted pulsar luminosity distribution.} 
\end{figure}



\section{The oldest X-ray supernovae}

A spin-off of our observational study \cite{sor08} was the discovery or recovery 
of three pre-1970 X-ray SNe, that is in the characteristic age range 
when the outgoing shock reaches the interstellar medium.
The oldest X-ray SN we have discovered is SN\,1941C, 
with $L_{\rm X} \approx 5 \times 10^{37}$ erg s$^{-1}$ 
in the $0.3$--$10$ keV band (Table 1). 
For SN\,1959D, we infer $L_{\rm X} \sim$ a few $10^{37}$ erg s$^{-1}$. 
For SN\,1968D, the emitted luminosity $L_{\rm X} \approx 2 \times 10^{38}$ erg s$^{-1}$, 
mostly thermal (i.e., from the shocked gas), but with an additional power-law 
component with photon index $\approx 2$ and $L_{\rm X} \approx 2 \times 10^{37}$ erg s$^{-1}$. 
In total, there are only 27 historical core-collapse SNe from 1900--1970, 
at distances $\la 15$ Mpc (for many other old SNe, the type is undetermined). 
Of them, 17 are so far undetected in X-rays, down to $\sim$ a few $10^{37}$ 
erg s$^{-1}$; 4 are detected at similar luminosities; 
the remaining 6 have not yet been observed by {\it Chandra} or {\it XMM-Newton}  
(apart from snapshots, too short to provide meaningful constraints). 


For our fitted luminosity of SN\,1941C, we infer a mass-loss rate 
from the progenitor star 
$\approx 5 \times 10^{-5} M_{\odot}$ yr$^{-1}$ at $\approx 55,000$ yr 
before the SN \cite{sor08}. For SN\,1968D, the inferred mass-loss rate is  
$\approx 8 \times 10^{-5} M_{\odot}$ yr$^{-1}$ at $\approx 30,000$ yr 
before the SN. 
The X-ray luminosity and mass-loss rate of SN\,1968D 
suggest that this SN is evolving to a remnant similar to Cas A. 
The similarity between SN\,1968D and Cas A was also noted from radio observations 
\cite{hym95}. The radio behaviour of SN\,1968D was also 
found to be similar to that of another SN in NGC\,6946, 1980K 
\cite{hym95}, which we have recovered in the {\it Chandra} 
observations from 2002--2004, at a luminosity $\approx 4 \times 10^{37}$ 
erg s$^{-1}$, in the $0.3$--$8$ keV band \cite{sor08}. 
The presence of a power-law-like emission component above 2 keV in
SN\,1968D suggests that there may be a contribution from a young
pulsar and its wind nebula; its luminosity $\approx 10^{37}$
erg s$^{-1}$ is comparable to the luminosity of Crab-like systems \cite{pos02}, 
and in particular the Crab itself and PSR\,B0540$-$69 \cite{ser04}.

\begin{figure}
\psfig{file=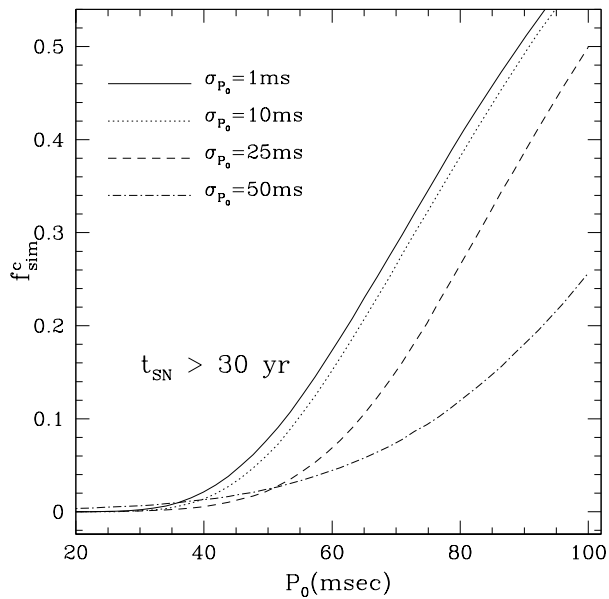,width=0.47\textwidth}
\caption{\footnotesize Fraction $f_{sim}^c$ of Monte Carlo realizations of the SN sample 
for which the predicted pulsar luminosities are consistent with the observed 
limits of the corresponding SNe. This is shown for different distributions
of the initial spin periods, described by Gaussians of mean $P_0$ and
dispersion $\sigma_{P_0}$. This sample includes only SNe older than 30 yrs.}
\end{figure}

\begin{table}
\renewcommand{\baselinestretch}{1.}
{\footnotesize
\caption{
Unambiguously-identified core-collapse SNe in nearby galaxies ($d \le 15$ Mpc) until 1970, 
and luminosity or upper limits to the {\it power-law component} of their 
X-ray emission (upper limit to the luminosity of a possible young pulsar). 
All values derived from {\it Chandra} data, except where noted.}
\begin{center}
\begin{tabular}{lccccr} 
%
\hline \hline
\multicolumn{1}{c}{SN ID} & \multicolumn{1}{c}{Type} &
\multicolumn{1}{c}{Galaxy} & \multicolumn{1}{c}{Distance} &
 \multicolumn{1}{c}{$L_{\rm po, 0.3-8}$} & \multicolumn{1}{r}{Notes}\\
  & & & \multicolumn{1}{c}{(Mpc)} & \multicolumn{1}{c}{(erg s$^{-1}$)} &\\
\hline \\[-6pt]
1909A	& II  & M101 & 7.4 &	 $< 5 \times  10^{36}$\\[2pt]
1917A	& II  & N6946 & 5.5 & 	 $< 2 \times  10^{36}$\\[2pt]
1921B	& II  & N3184 & 8.7 &	 $< 10^{37}$\\[2pt]
1923A   & IIP & M83  & 4.5 &     $< 3 \times 10^{36}$\\[2pt]
1937A   & IIP & N4157 & 15 &     $<3 \times 10^{37}$ & $(a)$\\[2pt]
1937F   & IIP & N3184 & 8.7 &    $<10^{37}$\\[2pt]
1940A   & IIL & N5907 & 13 &     $<2 \times 10^{37}$ & $(a)$\\[2pt]
1940B   & IIP & N4725 & 13 &     $<3 \times 10^{37}$\\[2pt]
1941A   & IIL & N4559 & 10 &     $<2 \times 10^{37}$\\[2pt]
1941C	& II  & N4136 & 10 &     $5 \times 10^{37}$ & $(b)$\\[2pt]
1948B	& IIP  & N6946 & 5.5 & 	 $< 2 \times  10^{36}$\\[2pt]
1951H	& II  & M101 & 7.4 &	 $< 5 \times  10^{36}$\\[2pt]
1954A   & Ib  & N4214  & 3.0 &   $< 5 \times 10^{35}$\\[2pt]
1959D	& IIL  & N7331 & 15&     $\sim$ a few $10^{37}$ & $(b)$\\[2pt]
1961U	& II  & N3938 & 12 &     ?\\[2pt]
1961V   & IIn  & N1058  & 9.1 &  $<3 \times 10^{37}$\\[2pt]
1962L	& Ic  & N1073 & 15 &     $<2 \times 10^{38}$\\[2pt]
1962M   & IIP  & N1313  & 4.0 &  $< 2 \times 10^{36}$\\[2pt]
1964H	& II  & N7292 & 15 &     ?\\[2pt]
1964L	& II  & N3938 & 12 &     ?\\[2pt]
1966J	&  Ib  &  N3198 &  12 &  ?\\[2pt]
1968D	& II  & N6946 & 5.5 & 	 $2 \times 10^{37}$ & $(b)$\\[2pt]
1968L   & IIP & M83  & 4.5 &     $< 3 \times 10^{36}$\\[2pt]
1969B   & IIP  & N3556  & 14 &   $< 3 \times 10^{37}$\\[2pt]
1969L   & IIP  & N1058  & 9.1 &  $<3 \times 10^{37}$\\[2pt]
1970A & II & IC3476 & 10 &       ?\\[2pt]
1970G   & IIL & M101 & 7.4 &	 $\approx 10^{37}$ & $(c)$\\[2pt]
\hline\\[-5pt]
\end{tabular}
\end{center}
\vspace{-0.2cm}
{\sc Notes}~--~$^{(a)}$From {\it XMM-Newton} data; 
$^{(b)}$ Soria \& Perna \shortcite{sor08}; 
$^{(c)}$ Immler \& Kuntz \shortcite{imm05}.
\vspace{-0.4cm}
}
\end{table}

\section{Constraints on birth parameters}

We started from the results of a recent large-scale radio pulsar survey 
\cite{arz02}: for a dipole approximation, 
the magnetic field has a log-normal distribution with a mean value
$\log B_0(\rm {Gauss}) = 12.35$, $\sigma_{B_0}=0.4$; 
the initial birth period distribution has a mean  $\log P_0 (\rm{s})= -2.3$, 
$\sigma_{P_0}>0.2$.
In order to test the resulting theoretical predictions for the pulsar distribution 
of X-ray luminosities, we performed Monte Carlo realizations 
of the compact object remnant population.
The fraction of SNe that leave behind a NS has been 
theoretically estimated by Heger et al. \shortcite{heg03} as $\approx 87$\% 
(subject to variability depending on stellar metallicity and initial mass function). 
If an object is a BH, we assigned it a luminosity $< 10^{35}$ erg s$^{-1}$. 
For the NSs, the birth period and magnetic field in our simulation were drawn 
from the Arzoumanian distribution, and were evolved to the current age. 
The corresponding X-ray luminosity was then drawn from a log-Gaussian distribution 
derived from the Possenti relation.
When comparing the predicted and observed X-ray luminosity distributions, 
we need to consider the possibility of obscuration.
However, we suggest (see discussion in Perna et al.~2008) that 
neutral absorption does not strongly affect our results, particularly 
in the $2$--$10$ keV band, because we considered 
only SNe older than 30 years (an analysis of all those older than 10 years 
gives the same statistical result); also, luminous X-ray pulsars would ionize 
the entire mass of the ejecta on a time-scale between a few years 
and a few tens of years.

We find (Fig.~1) that out of the 106 Monte Carlo realizations of the sample, 
none of them predicts pulsar luminosities compatible with the observed SN X-ray limits.
The predicted high-luminosity pulsars, corresponding to those with the shortest 
initial periods, are not observed. 
This suggest that the mean initial period of the pulsar population 
is slower than the millisecond periods derived from some population-synthesis 
studies in the radio \cite{arz02}. 
A number of other recent investigations 
have reached similar conclusions. For example, 
the population-synthesis studies of Faucher-Giguere \& Kaspi \shortcite{fau06} yielded 
a good fit to the data with a mean birth period of $0.3$ s, $\sigma =0.15$ s, 
for $\log B_0(\rm {Gauss}) = 12.35$. 
Similarly, the analysis by Ferrario \& Wickramasinghe \shortcite{fer06} yielded 
a mean period of $0.23$ s for a magnetic field of $10^{12}$ Gauss. 
We performed Monte Carlo simulations of the X-ray 
pulsar population using those birth parameters  
and found them to be consistent with our SNe X-ray limits.
We obtain a quantitative limit on the birth period by 
plotting (Fig.~2) the fraction of Monte Carlo simulations for which 
the luminosity of each pulsar is found below that of the corresponding SN 
(that is, the fraction of ``acceptable'' configurations), for different 
values of standard deviations. Mean periods shorter than $\approx 40$ ms are 
ruled out at the $90$\% confidence limit, for any assumed value 
of the width of the period distribution \cite{per08}. This is also consistent 
with studies of young Galactic SNRs \cite{sri84}.

A possible reason for the lack of luminous X-ray pulsars is a turnover 
in the efficiency of conversion of rotational energy into X-ray luminosity 
(relation between $\dot{E}_{\rm dot}$ and $L_{\rm X}$), at high values of $\dot{E}_{\rm dot}$.
Another systematic effect that might affect our results is that  
a fraction of NSs may be born with a non-active magnetosphere.
Similarly, the precise fraction of BHs versus NSs in the remnant population plays 
a role in our results. A larger fraction of BHs would alleviate our constraints 
on the initial spin periods, while a smaller fraction would make them tighter. 
Our results for the bulk of the pulsar population do not exclude that the subpopulation 
of magnetars could be born with very fast spins.
But their spin-down times would be very short, and the spin-down X-ray luminosity 
would decline within the first few months, during which the SN is still too optically thick 
to let the pulsar X-ray photons go through. Therefore, our analysis 
cannot place meaningful constraints on this class of objects.

Finally, our results have implications on the contribution of young X-ray pulsars 
to the population of ultraluminous X-ray sources observed in nearby galaxies. Earlier models  
predicted that a sizable fraction of that population could be made up of young, 
fast-rotating pulsars \cite{per04}. However, our study shows that 
the contribution from this component, although expected from the tail 
of the population, cannot be significant.

\section*{Acknowledgment}
RS is grateful to Prof.~Shuang-Nan Zhang for his generous 
hospitality at Tsinghua University, for his invitation to take part 
in this conference, and for his financial support.

\label{lastpage}

\clearpage

\end{document}

\label{lastpage}

\clearpage

\end{document}